\documentclass[aps,superscriptaddress,twocolumn,twoside,floatfix,prx,nofootinbib,a4paper]{revtex4-2}
\usepackage[export]{adjustbox} 
\usepackage{times}
\usepackage{amsmath,amssymb,amsthm,mathtools,thmtools,nameref}
\usepackage{comment}
\usepackage{enumerate}
\usepackage{float}
\usepackage{xcolor}
\usepackage[colorlinks=true,linkcolor=blue,citecolor=magenta,urlcolor=blue]{hyperref}
\usepackage{csvsimple}
\usepackage[capitalize]{cleveref}
\usepackage{braket}
\usepackage{enumitem}
\usepackage{placeins}
\usepackage{quantikz}

\tikzset{
  wgate-base/.style={
    draw=black,
    inner sep=2pt, minimum width=1.2em, minimum height=1.4em,
  },
  wtop/.style={
    wgate-base,
    path picture={
      \draw[decorate, decoration={zigzag, amplitude=1.2pt, segment length=3pt}]
        (path picture bounding box.north west) -- (path picture bounding box.north east);
    }
}}

\DeclareMathOperator{\Tr}{Tr}
\DeclareMathOperator{\polylog}{polylog}
\renewcommand{\O}{\mathcal O}

\newcommand{\id}{\mathbb{I}}
\newcommand{\prlsection}[1]{\textit{#1.}---}

\newcommand{\mat}[1]{\begin{pmatrix} #1 \end{pmatrix}}

\usepackage{tikz}
\usetikzlibrary{decorations.pathreplacing,calligraphy,decorations.markings,decorations.pathmorphing,math}

\definecolor{tensor}{rgb}{0.5,0.8,0.5}
\definecolor{isometry}{rgb}{0.8,0.8,1}

\newcommand{\targetmps}{
  	\begin{array}{c}
	\begin{tikzpicture}[scale=.2,thick]
        \draw[opacity=0.5, dash pattern=on 1pt off 1pt] (-5.6,0)--(-4.5,0);
        \draw[opacity=0.5] (-4.5,0)--(-4,0);
    	\draw[opacity=0.5] (-3,1)--(-3,2);
        \draw[opacity=0.5] (24,1)--(24,2);
        \draw[opacity=0.5] (25,0)--(25.5,0);
        \draw[opacity=0.5, dash pattern=on 1pt off 1pt] (26.6,0)--(25.5,0);
		\draw (-2,0)--(23,0);
		\draw (0,0)--(0,2);
		\draw (3,0)--(3,2);
		\draw (6,0)--(6,2);
		\draw (9,0)--(9,2);
		\draw (12,0)--(12,2);
        \draw (15,0)--(15,2);
		\draw (18,0)--(18,2);
		\draw (21,0)--(21,2);
        \filldraw[fill=tensor, opacity=0.5] (-4,-1) rectangle (-2,1);
		\filldraw[fill=tensor] (-1,-1) rectangle (1,1);
		\filldraw[fill=tensor] (+2,-1) rectangle (4,1);
		\filldraw[fill=tensor] (+5,-1) rectangle (7,1);
		\filldraw[fill=tensor] (+8,-1) rectangle (10,1);
		\filldraw[fill=tensor] (+11,-1) rectangle (13,1);
        \filldraw[fill=tensor] (+14,-1) rectangle (16,1);
		\filldraw[fill=tensor] (+17,-1) rectangle (19,1);
		\filldraw[fill=tensor] (+20,-1) rectangle (22,1);
        \filldraw[fill=tensor, opacity=0.5] (23,-1) rectangle (25,1);
		\draw (0,0) node {\scriptsize $A$};
		\draw (3,0) node {\scriptsize $A$};
		\draw (6,0) node {\scriptsize $A$};
		\draw (9,0) node {\scriptsize $A$};
		\draw (12,0) node {\scriptsize $A$};
        \draw (15,0) node {\scriptsize $A$};
		\draw (18,0) node {\scriptsize $A$};
		\draw (21,0) node {\scriptsize $A$};
        \draw[opacity=0.5] (-3,0) node {\scriptsize $A$};
		\draw[opacity=0.5] (24,0) node {\scriptsize $A$};
	\end{tikzpicture}
	\end{array}
}
\newcommand{\blockedmps}{
  	\begin{array}{c}
	\begin{tikzpicture}[scale=.2,thick, decoration={zigzag, segment length=1.1mm, amplitude=0.3mm}]
        \filldraw[fill=tensor, opacity=0.5]  decorate{(-4,-1) -- (-4,1)} -- (-2,1) -- (-2,-1) -- cycle;
        \filldraw[fill=tensor, opacity=0.5]  decorate{(25,-1) -- (25,1)} -- (23,1) -- (23,-1) -- cycle;
        \draw[opacity=0.5] (-3,1)--(-3,2);
        \draw[opacity=0.5] (24,1)--(24,2);

		\draw (-2,0)--(23,0);
		\draw (0,0)--(0,2);
		\draw (3,0)--(3,2);
		\draw (6,0)--(6,2);
		\draw (9,0)--(9,2);
		\draw (12,0)--(12,2);
        \draw (15,0)--(15,2);
		\draw (18,0)--(18,2);
		\draw (21,0)--(21,2);
		\filldraw[fill=tensor] (-1,-1) rectangle (10,1);
		\filldraw[fill=tensor] (+11,-1) rectangle (22,1);
		\draw (5,0) node {\scriptsize $(A)^4$};
		\draw (17,0) node {\scriptsize $(A)^4$};
	\end{tikzpicture}
	\end{array}
}
\newcommand{\decomposedmps}{
  	\begin{array}{c}
	\begin{tikzpicture}[scale=.2,thick, decoration={zigzag, segment length=1.1mm, amplitude=0.3mm}]
        \filldraw[fill=tensor, opacity=0.5]  decorate{(-4,-1) -- (-4,1)} -- (-2,1) -- (-2,-1) -- cycle;
        \filldraw[fill=tensor, opacity=0.5]  decorate{(25,-1) -- (25,1)} -- (23,1) -- (23,-1) -- cycle;
        \filldraw[fill=isometry, opacity=0.5]  decorate{(-4,2) -- (-4,4)} -- (-2,4) -- (-2,2) -- cycle;
        \filldraw[fill=isometry, opacity=0.5]  decorate{(25,2) -- (25,4)} -- (23,4) -- (23,2) -- cycle;
        \draw[opacity=0.5] (-3,1)--(-3,2);
        \draw[opacity=0.5] (24,1)--(24,2);
        \draw[opacity=0.5] (-3,4)--(-3,5);
        \draw[opacity=0.5] (24,4)--(24,5);

		\draw (-2,0)--(23,0);
		\draw (0,0)--(0,5);
		\draw (3,3)--(3,5);
		\draw (6,3)--(6,5);
		\draw (9,0)--(9,5);
		\draw (12,0)--(12,5);
        \draw (15,3)--(15,5);
		\draw (18,3)--(18,5);
		\draw (21,0)--(21,5);
		\filldraw[fill=tensor] (-1,-1) rectangle (10,1);
		\filldraw[fill=tensor] (+11,-1) rectangle (22,1);
        \filldraw[fill=isometry] (-1,2) rectangle (10,4);
		\filldraw[fill=isometry] (+11,2) rectangle (22,4);
		\draw (5,0) node {\scriptsize $P_4$};
		\draw (17,0) node {\scriptsize $P_4$};
        \draw (5,3) node {\scriptsize $V_4$};
		\draw (17,3) node {\scriptsize $V_4$};
	\end{tikzpicture}
	\end{array}
}
\newcommand{\twicedecomposedmps}{
  	\begin{array}{c}
	\begin{tikzpicture}[scale=.2,thick, decoration={zigzag, segment length=1.1mm, amplitude=0.3mm}]
        \filldraw[fill=tensor, opacity=0.5]  decorate{(-4,-1) -- (-4,1)} -- (-2,1) -- (-2,-1) -- cycle;
        \filldraw[fill=tensor, opacity=0.5]  decorate{(25,-1) -- (25,1)} -- (23,1) -- (23,-1) -- cycle;
        \filldraw[fill=isometry, opacity=0.5]  decorate{(-4,2) -- (-4,4)} -- (-2,4) -- (-2,2) -- cycle;
        \filldraw[fill=isometry, opacity=0.5]  decorate{(25,2) -- (25,4)} -- (23,4) -- (23,2) -- cycle;
        \filldraw[fill=isometry, opacity=0.5]  decorate{(-4,5) -- (-4,7)} -- (-2,7) -- (-2,5) -- cycle;
        \filldraw[fill=isometry, opacity=0.5]  decorate{(25,5) -- (25,7)} -- (23,7) -- (23,5) -- cycle;
        \draw[opacity=0.5] (-3,1)--(-3,2);
        \draw[opacity=0.5] (24,1)--(24,2);
        \draw[opacity=0.5] (-3,4)--(-3,5);
        \draw[opacity=0.5] (24,4)--(24,5);
        \draw[opacity=0.5] (-3,7)--(-3,8);
        \draw[opacity=0.5] (24,7)--(24,8);

		\draw (-2,0)--(23,0);
		\draw (0,0)--(0,8);
		\draw (3,6)--(3,8);
		\draw (6,6)--(6,8);
		\draw (9,3)--(9,8);
		\draw (12,3)--(12,8);
        \draw (15,6)--(15,8);
		\draw (18,6)--(18,8);
		\draw (21,0)--(21,8);
		\filldraw[fill=tensor] (-1,-1) rectangle (22,1);
		\filldraw[fill=isometry] (-1,2) rectangle (22,4);
        \filldraw[fill=isometry] (-1,5) rectangle (10,7);
		\filldraw[fill=isometry] (+11,5) rectangle (22,7);
		\draw (11,0) node {\scriptsize $P_8$};
		\draw (11,3) node {\scriptsize $V_8$};
        \draw (5,6) node {\scriptsize $V_4$};
		\draw (17,6) node {\scriptsize $V_4$};
	\end{tikzpicture}
	\end{array}
}
\newcommand{\alternativeapproximatedmps}{
  	\begin{array}{c}
	\begin{tikzpicture}[scale=.2,thick, decoration={zigzag, segment length=1.1mm, amplitude=0.3mm}]
        \filldraw[fill=isometry, opacity=0.5]  decorate{(-4,2) -- (-4,4)} -- (-2,4) -- (-2,2) -- cycle;
        \filldraw[fill=isometry, opacity=0.5]  decorate{(25,2) -- (25,4)} -- (23,4) -- (23,2) -- cycle;
        \filldraw[fill=isometry, opacity=0.5]  decorate{(-4,5) -- (-4,7)} -- (-2,7) -- (-2,5) -- cycle;
        \filldraw[fill=isometry, opacity=0.5]  decorate{(25,5) -- (25,7)} -- (23,7) -- (23,5) -- cycle;
        \draw[opacity=0.5] (-3,0)--(-3,2);
        \draw[opacity=0.5] (24,0)--(24,2);
        \draw[opacity=0.5] (-3,4)--(-3,5);
        \draw[opacity=0.5] (24,4)--(24,5);
        \draw[opacity=0.5] (-3,7)--(-3,8);
        \draw[opacity=0.5] (24,7)--(24,8);

		\draw (-3,0)--(0,0);
        \draw[opacity=0.5] (21,0)--(21.4,0);        
        \draw[opacity=0.5] (23.6,0)--(24,0);
		\draw (0,0)--(0,8);
		\draw (3,6)--(3,8);
		\draw (6,6)--(6,8);
		\draw (9,3)--(9,8);
		\draw (12,3)--(12,8);
        \draw (15,6)--(15,8);
		\draw (18,6)--(18,8);
		\draw (21,0)--(21,8);
        \filldraw[color=black, fill=white, thick](-1.5, 0) circle (1.1);
        \filldraw[color=black, fill=white, thick, opacity=0.5](22.5, 0) circle (1.1);
		\filldraw[fill=isometry] (-1,2) rectangle (22,4);
        \filldraw[fill=isometry] (-1,5) rectangle (10,7);
		\filldraw[fill=isometry] (+11,5) rectangle (22,7);
        \draw (-1.5,0) node {\scriptsize $\omega$};
        \draw[opacity=0.5] (22.5,0) node {\scriptsize $\omega$};
		\draw (11,3) node {\scriptsize $V_8$};
        \draw (5,6) node {\scriptsize $V_4$};
		\draw (17,6) node {\scriptsize $V_4$};
	\end{tikzpicture}
	\end{array}
}

\newcommand{\ibmquantum}{IBM Quantum, IBM Research Europe - Zurich, 8803 R\"uschlikon, Switzerland}
\newcommand{\ethz}{Institute for Theoretical Physics, ETH Z\"urich, 8093 Zurich, Switzerland}

\begin{document}

\title{Renormalization-group-based preparation of matrix product states on up to 80 qubits}

\author{Moritz Scheer}
\affiliation{\ibmquantum}
\affiliation{\ethz}
\author{Alberto Baiardi}
\affiliation{\ibmquantum}
\author{Elisa B\"aumer Marty}
\affiliation{\ibmquantum}
\author{Zhi-Yuan Wei}
\affiliation{Joint Quantum Institute and Joint Center for Quantum Information and Computer Science,
NIST/University of Maryland, College Park, Maryland 20742, USA}
\author{Daniel Malz}
\affiliation{Department of Mathematical Sciences, University of Copenhagen, 2100 Copenhagen, Denmark}

\date{\today}

\begin{abstract}
A key challenge for quantum computers is the efficient preparation of many-body entangled states across many qubits.
In this work, we demonstrate the preparation of matrix product states (MPS) using a renormalization-group(RG)-based quantum algorithm~\cite{Malz_2024} on superconducting quantum hardware.
Compared to sequential generation, it has been shown that the RG-based protocol asymptotically prepares short-range correlated MPS with an exponentially shallower circuit depth (when scaling system size), but it is not yet clear for which system sizes it starts to convey an advantage.
We thus apply this algorithm to prepare a class of MPS exhibiting a phase transition between a symmetry-protected topological (SPT) and a trivial phase for systems of up to 80 qubits.
We find that the reduced depth of the RG-based circuits makes them more resilient to noise, and that they generally outperform the sequential circuits for large systems, as we showcase by measuring string-order-like local expectation values and energy densities.
We thus demonstrate that the RG-based protocol enables large-scale preparation of MPS and, in particular, SPT-ordered states beyond the fixed point.
\end{abstract}

\maketitle

\prlsection{Introduction} The preparation of quantum states is a crucial subroutine in quantum computing, particularly when simulating quantum systems. In such cases, some initial state, often the ground state of a Hamiltonian of interest, must be prepared. Preparing quantum states is exponentially complex in general. Hence, the identification of efficient preparation algorithms for interesting classes of quantum states is of great importance. Matrix product states (MPS)~\cite{Fannes_1992,Hastings_2007,McCulloch_2007} are a prominent class of relevant quantum states, containing the ground states of one-dimensional, gapped Hamiltonians and paradigmatic quantum states such as the GHZ state~\cite{Greenberger1989}, W state~\cite{Dur_2000}, cluster state~\cite{Briegel_2001}, and the AKLT state~\cite{Affleck_1987}.

Several MPS preparation algorithms have been proposed, which can be broadly classified as heuristic or with asymptotic accuracy guarantees. Heuristic protocols usually rely on the variational approach, where a parametrized ansatz circuit is classically optimized to maximize the fidelity of the circuit output state with the target MPS~\cite{Ran2020_EncodingMPS,Rudolph2023_MPS,iqbal2022preentangling,wei2025state2}. A key limitation of these algorithms, however, is that controlling the approximation error associated with the ansatz choice is non-trivial. Moreover, optimizing a given ansatz can become hard for large MPS.
Rigorous protocols prepare MPS either exactly or with a controllable approximation error. The most straightforward preparation protocol is \textit{sequential} preparation, which encodes the MPS as a ladder-like quantum circuit~\cite{Schoen_2005}.
This protocol, which has been used in recent quantum algorithms and experiments~\cite{smith2022crossing,Potter2022_qMPO,Wall2024_qMPS-Dynamics,Anselme2024_MPS-Trotter,Schuhmacher2025_hTTN}, prepares the target MPS exactly in a circuit depth that scales linearly in system size.

Recent research has shown that in many circumstances, MPS can be prepared much faster. For instance, \textit{fusion}-based approaches, leveraging dynamic circuits, which perform mid-circuit measurements followed by classical processing and feed-forward operations, prepare particular MPS in constant depth~\cite{Smith2023_AKLT-Fusion,Sahay2024_FiniteDepth-TNS,Stephen2024_MPS-Fusion-Characterization}.
This, however, works only for specific classes of MPS~\cite{Styliaris2024_Fusion-PEPS}.
A more general method is adiabatic preparation, which can be used to prepare all gapped MPS in depth $\O(\polylog(n/\epsilon))$~\cite{Ge2016,zyadi}.

The arguably fastest general MPS preparation algorithm is based on the \textit{renormalization} group (RG) transformation~\cite{Malz_2024}, combining rigorous performance guarantees, broad applicability to arbitrary states, and shallow circuit depth. While only approximately encoding the target MPS up to an error $\epsilon$ in the fidelity, it prepares \emph{any} short-range correlated MPS with a unitary, local circuit in depth $\mathcal{O}(\log (n /\epsilon))$.
Dynamic circuits extend its applicability to long-range correlated MPS and further reduce the circuit depth scaling to $\mathcal{O}(\log \log (n/\epsilon))$.
While this asymptotically exponential reduction in circuit depth has been rigorously established, it is not yet clear for which  system sizes and noise levels the RG-based protocol becomes advantageous in practice.

In this work, we compare practical quantum-hardware implementations of both the sequential and the RG-based protocols. We apply them to prepare the ground states of a one-dimensional spin-$1/2$ model for a range of parameters that include a phase transition between its trivial and symmetry-protected topological phase~\cite{Wolf_2006} on up to 80 qubits using the IBM Heron quantum processor. The parameters are chosen such that the ground state along the whole path is a bond dimension $D=2$ MPS.
We validate the prepared states by measuring two string-order parameters and the energy density.
As our main result, we find that the RG-based preparation protocol already yields a practical advantage on currently available quantum hardware beyond a certain system size. The preparation shows significant improvements in system size and fidelity with respect to earlier results on 9 qubits~\cite{smith2022crossing}.
Thus, our experiment represents the largest demonstration to date of the preparation of states in an SPT ordered phase away from the fixed point.
As a side result, we also uncover that the required circuit depth in the RG-based protocol can be reduced significantly by choosing an appropriate MPS gauge.

\prlsection{Target MPS}We consider the one-dimensional spin-$1/2$ system with Hamiltonian 
\begin{equation}
    H = \sum_{i=1}^{n} \left( -g_{zz}  Z_i Z_{i+1} - g_x   X_i + g_{zxz}  Z_i X_{i+1} Z_{i+2} \right),
    \label{equ:mps_family_hamiltonian}
\end{equation}
where $n$ is the number of spins, $X$, $Y$, and $Z$ correspond to Pauli matrices and the indices are defined modulo $n$, giving periodic boundary conditions. This system, first defined in Ref.~\cite{Wolf_2006}, has three distinct phases: a trivial, a symmetry-broken, and a symmetry-protected topological (SPT) phase.
There exists a one-dimensional path in phase space along which the ground state corresponds exactly to a translation-invariant (TI) MPS of bond dimension $D=2$.
It is given with tuning parameter $g\in[-1,1]$ by~\cite{Wolf_2006} 
\begin{equation}
    g_{zz}  = 2(1-g^2),\quad
    g_x     = (g + 1)^2,\quad
    g_{zxz} = (1 - g)^2.
    \label{eq:G_path}
\end{equation}
The path connects the paramagnetic Hamiltonian $H_X = -4 \sum_{i=1}^n X_i$ at $g=1$ in the trivial phase (for all $g > 0$) with the cluster Hamiltonian $H_{ZXZ}= 4 \sum_{i=1}^n Z_i X_{i+1} Z_{i+2}$ at $g=-1$ in the SPT phase (for all $g<0$). The phase transition occurs at the tricritical point ($g=0$). The correlation length $\xi$ of the ground states is 0 for $|g|=1$ and monotonically increases for $|g|\to 0$ according to $\xi=|\ln[(1-g)/(1+g)]|^{-1}$. The ground states at $g=1$, $g=0$, and $g=-1$ correspond to $|+\rangle^{\otimes N}$, the GHZ state, and the cluster state, respectively. 
Combining simplicity, nontrivial physics, and tunable correlation length, this family of states has emerged as a popular benchmark for state preparation algorithms~\cite{smith2022crossing,Smith2024mps,zyadi,wei2025state2}.

The TI MPS ansatz for a chain of $n$ spin-$1/2$ sites with bond dimension $D$ reads (suppressing a normalization factor)
\begin{equation}
    \ket{\Psi} \propto \sum_{i_1, ..., i_n=0}^{1} 
    \Tr \left( A^{(i_1)} A^{(i_2)} \cdots A^{(i_n)} \right)
    \ket{i_1 i_2 \cdots i_n},
    \label{eq:mps}
\end{equation}
where $A^{(i)}$ are $D\times D$ matrices. The matrices for the ground state along the path of \cref{eq:G_path} are given as~\cite{Wolf_2006,smith2022crossing}
\begin{equation}
    A^0 = \frac{1}{\sqrt{1+|g|}}\mat{0&0\\\sqrt g&1},\qquad A^1=XA^0X. 
	\label{eq:A}
\end{equation}

The phase transition between the trivial and SPT phase can be detected by measuring the two nonlocal observables $S^{\mathbb{I}}  = I_1 X_2 X_3 \cdots X_{n-1} I_n$ and $S^{ZY} = Z_1 Y_2 X_3 \cdots X_{n-2} Y_{n-1} Z_n$~\cite{smith2022crossing} corresponding to the string-order parameters $\mathcal{S}^\id = \langle S^\id \rangle$ and $\mathcal{S}^{ZY}=\langle S^{ZY} \rangle$. The former is non-zero in the trivial phase, while the latter is non-zero in the SPT phase. Their nonlocal nature is, however, detrimental to their measurement on noisy quantum hardware, since the readout fidelity reduces exponentially in the weight of the measured Pauli observable. Fortunately, there exist local observables $S_a^{\mathbb{I}}$ and $S_a^{ZY}$ defined as
\begin{align}    
    S_a^{\mathbb{I}} &= \frac{1}{n} \sum_{q=1}^n I_{q} X_{q+1} X_{q+2} X_{q+3} X_{q+4} X_{q+5} I_{q+6} \\
    S_a^{ZY} &= \frac{1}{n} \sum_{q=1}^n Z_{q} Y_{q+1} X_{q+2} Y_{q+3} Z_{q+4}.
\end{align}
that, on this particular family of states, yield the same expectation values~\cite{smith2022crossing} as $S^\id$ and $S^{ZY}$, respectively. (As above, indices are defined modulo $n$.) Note that we chose $S_a^{\mathbb{I}}$ to have the same Pauli support as $S_a^{ZY}$.

\prlsection{RG-based preparation protocol}We prepare the states discussed above with the renormalization-group(RG)-based preparation protocol proposed in Ref.~\cite{Malz_2024}, which we briefly review below. Graphically, we represent a part of the TI target MPS as
\begin{equation}
	\ket\Psi\propto\targetmps.
	\label{eq:target_mps}
\end{equation}
The RG-based protocol starts by blocking a certain number of sites $q_0$ such that $d^{q_0} \geq D^2$, where $d$ and $D$ denote the physical and bond dimension of the MPS, respectively. In our case with $d=D=2$, we choose $q_0=4$, transforming \cref{eq:target_mps} into
\begin{equation}
	\blockedmps.
	\label{eq:blocked_mps}
\end{equation}
Next, we apply the polar decomposition, which factorizes any matrix $M \in \mathbb{C}^{m \times n}$ with $m \geq n$ into an isometry $V \in \mathbb{C}^{m \times n}$ and a positive semidefinite matrix $P \in \mathbb{C}^{n \times n}$, on every block $(A)^4$ (where both the physical and virtual indices are grouped together during the decomposition).
This transforms \cref{eq:blocked_mps} into
\begin{equation}
	\decomposedmps,
	\label{eq:decomposed_mps}
\end{equation}
where $V_4$ and $P_4$ are the isometry and positive semidefinite matrix, respectively. This blocking-decomposition-step corresponds to an RG transformation of the MPS~\cite{Verstraete2005}. In the second step, two neighboring positive semidefinite tensors are blocked and decomposed in the same manner, transforming \cref{eq:decomposed_mps} further into
\begin{equation}
	\twicedecomposedmps.
	\label{eq:twice_decomposed_mps}
\end{equation}
By repeating these RG steps several times, the positive semidefinite tensor converges to a limit $P_\infty$, which corresponds to the fixed point of the RG transformation. In the RG-based preparation protocol, this process is terminated after a small number of steps and the final positive semidefinite tensor is replaced by its fixed point $P_\infty$. This replacement results in a controllable, small approximation error (see later).

For short-range correlated TI MPS, such as the states we prepare in this work, the fixed point corresponds to a tensor product of nearest-neighbor entangled pairs, $\ket{\omega} = (\mathbb{I} \otimes \sqrt{\rho}) \sum_{i=1}^D \ket{ii}$, where the matrix $\rho$ is a reshaped version of the leading right eigenvector of the fixed point $P_\infty$.
Here, we have $\ket\omega\propto(\ket{00}+\ket{11}+\sqrt g(\ket{01}+\ket{10}))$ for $g>0$ and $\ket\omega\propto\ket{00}+\ket{11}$ for $g\leq0$.
If we have blocked a total of $q$ sites in the RG steps, we replace the remaining positive semidefinite $P_q$ by $\ket{\omega}^{\otimes(n/q)}$. Applied to \cref{eq:twice_decomposed_mps} (with $q=8$), this yields 

\begin{equation}
    \ket{\phi_\mathrm{approx}}
    =\alternativeapproximatedmps.
	\label{eq:approximated_mps}
\end{equation}

\Cref{eq:approximated_mps} allows a straightforward implementation as a quantum circuit. First, the fixed point is prepared in constant depth, as all $\ket{\omega}$ can be prepared in parallel. After that, the isometries reintroduce the local correlations of the approximated target MPS. The isometries are extended to unitaries and then implemented as gates of the quantum circuit. All isometries have constant size, in our case $16 \times 4$. Due to their nonlocal nature, however, SWAP-gate ladders are required to implement them on platforms with a local connectivity.
This results in an overall circuit depth that scales linearly with block size $q$. Since we need $q\sim\log (n)$ to achieve a small approximation error for all $n$, the overall circuit depth of the RG-based protocol is $\mathcal{O}(\log (n))$ when accounting for local connectivity.

\prlsection{Approximation error}The error $\epsilon=1-|\langle \Psi| \phi_\mathrm{approx}\rangle|$ resulting from the fixed point approximation depends on the correlation length of the state $\xi = -\ln(1/\lambda)$ (where $\lambda$ is the subleading eigenvalue of the MPS transfer matrix), the system size $n$, 
and the blocking number $q$, and obeys $\epsilon = \mathcal{O}\left( (n/q) e^{- \gamma q / \xi} \right)$ for any constant $\gamma<1/2$~\cite{Malz_2024}. 
Thus, a larger $q$ gives a smaller approximation error, but also increases the circuit depth, which incurs more hardware noise. Consequently, when implementing the RG-based protocol on a noisy device, the $q$ value yielding the most accurate state preparation results from a trade-off between these two error sources.
In this work, we prepare states with variable correlation length depending on $g$. We therefore chose a suitable $q$ for each value of $g$ individually: we set $q=4$ for $g \in \{ \pm0.3, \pm0.5, \pm0.7, \pm0.9 \}$ and $q=8$ for the two states closest to the phase transition ($g=\pm0.1$). We do so for all $n \in \{16,32,48,64,80\}$ (and thereby accept a slight increase in $\epsilon$ with system size). Overall, this results in a fixed point approximation error of at most $7.5\%$ (see \cref{fig:fixed_point_approximation_error} in Appendix).

\prlsection{Implementation on hardware}To execute the RG-based preparation protocol on a quantum processor, the high-level circuit corresponding to \cref{eq:approximated_mps} must be transpiled to meet the hardware constraints of the target device.
On hardware with a limited coherence time and noisy gates, the transpiled circuit must be as shallow and sparse as possible, as the noise level at the circuit output increases with the circuit native two-qubit gate depth and count.

In our case, the implementation of the fixed point preparation is straightforward, as any two-qubit state can be prepared with a single CNOT gate \cite{Iten_2016}. The challenging part lies in the implementation of the isometries of dimension $16 \times 4$, which are embedded in four-qubit unitary gates. Ref.~\cite{Iten_2016} provides a transpilation method yielding 54 CNOT gates (assuming full connectivity between all 4 qubits). Fortunately, the symmetries of the target MPS are reflected in the isometries, which allows us to map the $16 \times 4$ isometries to a direct sum of two $8 \times 2$ isometries. With this insight and the $8 \times 2$ isometry transpilation of Ref.~\cite{Iten_2016}, the overall isometry can be transpiled (still assuming full connectivity between all 4 qubits) with only 13 CNOT gates. This reduction depends on the gauge of the MPS, and thus highlights the importance of optimizing the gauge choice in RG-based preparation algorithms. More details on the observed structure and this implementation can be found in \cref{app:transpilation}. Furthermore, we optimized the mapping of the virtual qubits of the quantum circuit to the physical qubits of the device, as explained in \cref{app:mapping}. Altogether, the native two-qubit gate depth of the transpiled circuit is 28 (27) for circuits with $q=4$ and 68 (82) for circuits with $q=8$ for $g>0$ ($g<0$).

\begin{figure*}[t]
    \centering
    \includegraphics[width=\linewidth, trim={0bp 20bp 0bp 0bp}]{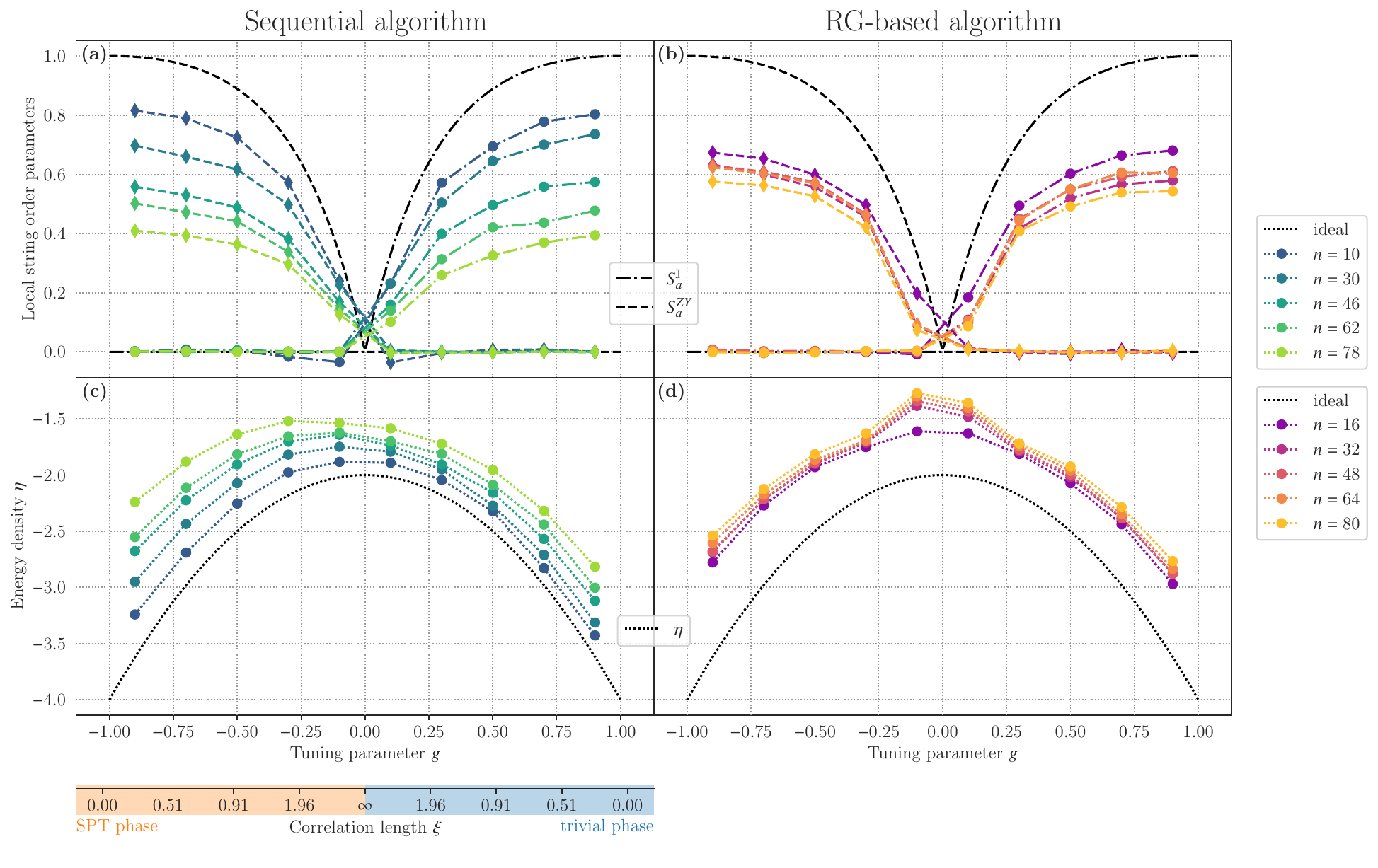}
    \caption{Experimental results comparing the local string-order parameters $\mathcal{S}_a^\id$ and $\mathcal{S}_a^{ZY}$ $\mathbf{(a,b)}$ and the energy density $\eta\coloneqq\langle H\rangle/n$ $\mathbf{(c,d)}$ measured on states prepared sequentially [\cref{eq:seq-circ}, $\mathbf{(a,c)}$] or with the RG-based algorithm [\cref{eq:approximated_mps}, $\mathbf{(b,d)}$], as a function of the tuning parameter $g$ and evaluated for varying number of qubits $n$. The (statistical) error bars in this plot are smaller than the plot markers. For reference, we also include an additional scale showing the correlation length $\xi$ corresponding to the parameter $g$. } 
    \label{fig:expresults}
\end{figure*}

\begin{figure*}[t]
    \centering
    \includegraphics[width=\linewidth]{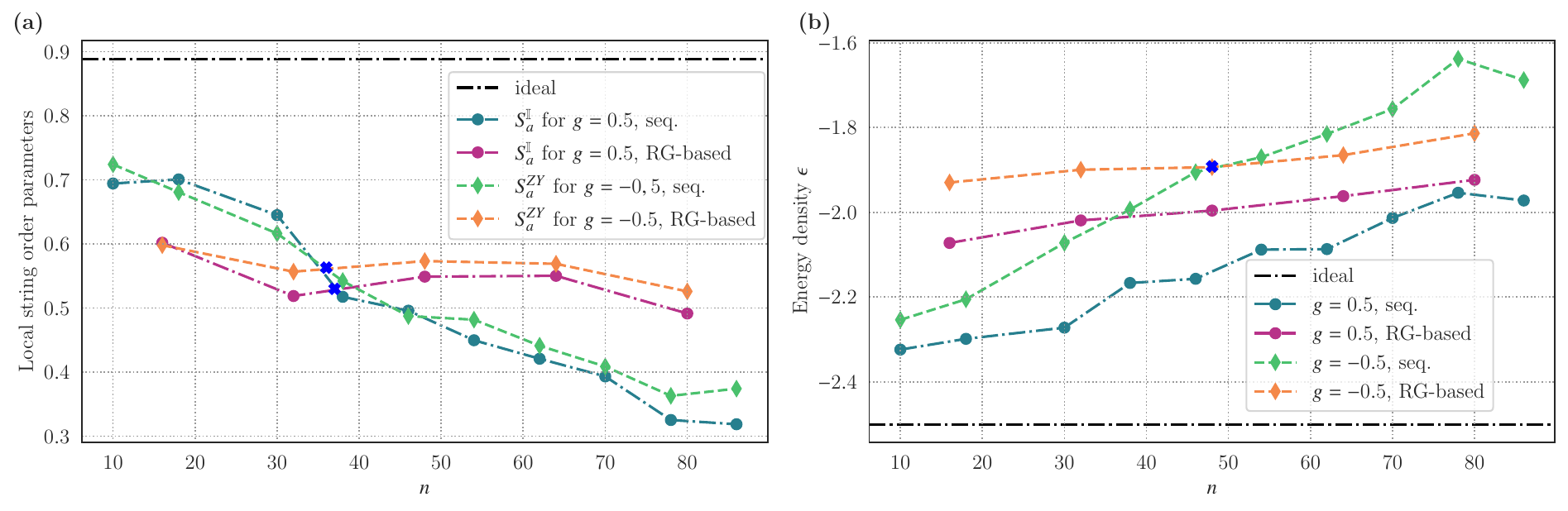}
    \caption{$\mathbf{(a)}$ and $\mathbf{(b)}$ show the local string-order parameters $\mathcal{S}^\id_a$ and $\mathcal{S}^{ZY}_a$ and the energy density $\eta$, respectively, as a function of the number of qubits $n$, evaluated for tuning parameters $g=0.5$ and $g=-0.5$. The blue x-markers denote the crossover point of the sequential and the RG-based preparation. Each plot also shows the ideal value.}
    \label{fig:expresults2}
\end{figure*}

\prlsection{Results}We prepare ground states of the Hamiltonian $H$ of \cref{equ:mps_family_hamiltonian} in its trivial and SPT phase on IBM's state-of-the-art Heron quantum processors. Precisely, we prepare the ground states along the path of \cref{eq:G_path}, where the ground states correspond to MPS of bond dimension $D=2$, for values of $g \in \{ \pm0.1, \pm0.3, \pm0.5, \pm0.7, \pm0.9 \}$ using the RG-based preparation protocol~\cite{Malz_2024}. We evaluate the accuracy of the preparation by measuring the string-order parameters $\mathcal{S}_a^\mathbb{I}$ and $\mathcal{S}_a^{ZY}$ and the energy density $\eta\coloneqq\langle H\rangle/n$, and compare it to the sequential preparation protocol (see \cref{app:sequential-protocol}). Since the RG-based protocol allows to prepare MPS with an asymptotically exponentially smaller depth compared to the sequential algorithm, it is expected to provide a higher accuracy above a certain system size when implemented on current noisy hardware. With our experiments, we assess whether this crossover point is observed in practice.

We ran all circuits using Qiskit~\cite{qiskit2024} on the \texttt{ibm\_fez} device without applying any error suppression or error mitigation techniques, as these compiler-level optimizations can make it more difficult to observe clean scaling. We report our results in \cref{fig:expresults}. Subplots (a) and (c) display the string-order parameters and energy density obtained with the sequential preparation protocol. The results show that while for small $n$, sequential preparation gives superior results, its accuracy quickly decreases with $n$, due to the linear growth of the depth of the corresponding circuit. Subplots (b) and (d) show the analogous results for the RG-based protocol. In this case, the accuracy remains nearly constant across all system sizes, as expected from its constant circuit depth. The small accuracy decrease can be ascribed to the non-uniform qubit quality of the \texttt{ibm\_fez} device. Since for each value of $n$, we mapped the circuit to the highest-quality ring of qubits, the average quality is expected to decrease for increasing $n$. Notice the bump in the energy density and the steep decrease of the string-order parameters at $g=\pm0.1$. This is due to the higher blocking number of $q=8$ required here to maintain a small fixed point approximation error despite the rapid increase of the ground state correlation lengths close to the phase transition. For all other states, satisfactory approximations were obtained already for $q=4$. Furthermore, observe the slightly worse accuracy of the energy density for negative values of $g$. We attribute this to the larger measurement error in determining the expectation value of the weight-3 Pauli terms in the Hamiltonian, which contribute more strongly to the energy density in this regime.

Comparing the results of sequential and RG-based preparation, we observe that while the former exhibits a higher accuracy for small $n$, the latter outperforms it for larger states for all values of $g$, except $g=\pm0.1$, where the correlation length is largest. \Cref{fig:expresults2} (a) and (b) depict the change in accuracy over the system size $n$ for the ground states at $g=\pm0.5$. For the two string-order parameters $\mathcal{S}_a^\id$ (only shown for $g=0.5$ since zero in SPT phase corresponding to $g<0$) and $\mathcal{S}_a^{ZY}$ (only shown for $g=-0.5$ since zero for $g>0$), we observe the crossover point at $n\approx37$ and $n\approx 35$, respectively. For the energy density, the crossover point is only observed for $g=-0.5$ at $n\approx48$, while the two curves do not cross for $g=0.5$ for the system sizes considered here. In these plots, we attribute the fluctuating nature of the curves to the compiler assigning different sets of qubits for different system sizes. Nonetheless, a smaller slope of the RG-based preparation protocol is clearly visible, indicating the superior asymptotic scaling of the protocol with system size.

\prlsection{Conclusion}We applied the RG-based preparation protocol~\cite{Malz_2024} to prepare bond dimension 2 MPS with a range of correlation lengths, representing ground states of a one-dimensional spin-$1/2$ Hamiltonian in both its trivial and symmetry-protected topological phase. 
Our results show that the asymptotic exponential advantage of the RG-based implementation over sequential circuits translates to a practical advantage already on current state-of-the-art quantum hardware. Moreover, our experiments serve as the largest demonstration to date of the preparation of states in an SPT phase away from the fixed point.

An interesting direction for future research is to explore further improvements that may be obtained using dynamic circuits, as well as to compare the RG-based preparation with different protocols, such as parallel-sequential circuits~\cite{wei2025state2} or variational approaches~\cite{Jaderberg2025_VariationalPreparationNormalMPS}.
Finally, it will be interesting to test the RG-based algorithm for larger bond dimension or inhomogeneous MPS.

\begin{acknowledgments}
We thank Maika Takita and Marcel Hinsche for fruitful discussions and support. This research was supported by the project RESQUE (Rethinking Quantum Simulations in the Quantum Utility Era, grant number 20QU-1\_225229), funded by the Swiss National Science Foundation. ZYW~acknowledges support from the U.S.~Department of Energy, Office of Science, Accelerated Research in Quantum Computing, Fundamental Algorithmic Research toward Quantum Utility (FAR-Qu). D.M.\ acknowledges financial support from the Novo Nordisk Foundation under grant numbers NNF22OC0071934 and NNF20OC0059939.
\end{acknowledgments}

\appendix
\section{Sequential preparation of states with PBC}
\label{app:sequential-protocol}

Matrix product states (MPS) with open boundary conditions (OBC) with physical dimension $d$ and bond dimension $D$ can immediately be written as a sequential circuit of gates acting on pairs neighboring $1+\log_d(D)$ qudits~\cite{Schoen_2005}. This map is not as immediate for MPS with periodic boundary conditions (PBC). In principle, they can always be written as MPS with OBC of bond dimension $D^2$, but this can incur significant overhead in circuit depth. The RG preparation does not have this limitation. 

To make the comparison fair, we use the following optimized sequential preparation procedure. We first prepare the OBC state in the sequential protocol and then perform one step of postselection to fuse the ends, which returns the desired state (with probability at least $D^2$). We further halve the depth of the sequential protocol by using a mixed canonical form, which allows one to write the state as two staircase circuits propagating outward from the middle.

In our specific case, we can rewrite the MPS as
\begin{equation}
\ket{\Psi}=
\begin{array}{c}
    \begin{tikzpicture}[scale=.35,thick,decoration={
                markings, mark=at position 0.8 with {\arrow{>}}}]
        \def\Nx{3}
        \tikzmath{\N=int(\Nx-1);}
        \foreach \x in {0,...,\N}{
            \begin{scope}[shift={(2*\x+1, 0)}]
                \draw[postaction={decorate}] (1/2, 0) -- (3/2, 0);
                \filldraw[fill=tensor] (-1/2,-1/2) rectangle (1/2,1/2);
                \draw[postaction={decorate}] (0, .5) -- (0, 1.2);
    		\draw (0,0) node {\scriptsize $A$};
            \end{scope}
            \begin{scope}[shift={(-2*\x-1, 0)}]
                \draw[postaction={decorate}] (-1/2, 0) -- (-3/2, 0);
                \filldraw[fill=tensor] (-1/2,-1/2) rectangle (1/2,1/2);
                \draw[postaction={decorate}] (0, .5) -- (0, 1.2);
    		\draw (0,0) node {\scriptsize $\tilde A$};
            \end{scope}
        }
        \draw[postaction={decorate}] (0, 0) -- ( 1/2, 0);
        \draw[postaction={decorate}] (0, 0) -- (-1/2, 0);
        \draw (-7.2,0)--(-7.5,0)--(-7.5,-.8)--(7.5,-.8)--(7.5,0)--(7.2,0);
        \draw[dash pattern=on 1pt off 1pt] (-7.5,0)--(-6.5,0);
        \draw[dash pattern=on 1pt off 1pt] ( 7.5,0)--( 6.5,0);
    \end{tikzpicture}
\end{array}
\label{eq:seq-circ}
\end{equation}
where $A^i_{\alpha\beta}$ is the tensor given in the main text \cref{eq:A}, and $\tilde A^i_{\alpha\beta}\coloneqq A^i_{\beta\alpha}$. 
In \cref{eq:seq-circ} we have added arrows to indicate that the tensors $A$ and $\tilde A$ are isometries from the Hilbert space of the incoming arrows to the joint Hilbert space of the outgoing arrows. This means that \cref{eq:seq-circ} is directly implementable as a quantum circuit, except for the boundary conditions, where the output of the leftmost and rightmost isometries is forced to be equal.
This last step is done by measuring in the Bell basis and postselecting on the state $\ket{\Phi_+}\propto\ket{00}+\ket{11}$.
In the middle, where $A$ and $\tilde A$ meet, one has to first prepare a Bell state $\ket{\Phi_+}$ between right and left qubit before applying the isometry circuits.
Note that this protocol can be generalized to arbitrary PBC MPS, but in general (i) bringing the MPS into mixed canonical form changes the tensors $A$ and $\tilde A$, (ii) there is a general entangled state in the middle (at the position of $\leftrightarrow$), not necessarily a Bell state.

\section{Transpilation details}
\label{app:expdetails}
\subsection{Implementation of isometries}
\label{app:transpilation}

The isometries of \cref{eq:approximated_mps} have dimension $16 \times 4$ for physical and bond dimension $d = D = 2$, corresponding to four-qubit unitary gates. Ref.~\cite{Iten_2016} provides an implementation of such isometries using 54 CNOT gates. For the isometries in this work, we discovered a much shallower implementation.
The rank-3 tensor of \cref{eq:A} has the following two properties:
\begin{subequations}
    \begin{align}
    \label{eq:A_prop1}
        (i)&\quad
        \begin{array}{c}
            \begin{tikzpicture}[scale=.2,thick]
                \draw (0,-3) node {};
                \draw (-2,0)--(2,0);
                \draw (0,0)--(0,2);
                \filldraw[fill=tensor] (-1,-1) rectangle (1,1);
                \draw (0,0) node {\scriptsize $A$};
                \draw (-3,0) node {\scriptsize $j$};
                \draw (0,3) node {\scriptsize $i$};                  
            \end{tikzpicture}
        \end{array}
        = 0 \qquad \forall \  i \neq j,\\
        (ii)&\quad
        \begin{array}{c}
            \begin{tikzpicture}[scale=.2,thick]
                \draw (0,-3) node {};
                \draw[white] (0,0)--(0,5);
                \draw (-2,0)--(2,0);
                \draw (0,0)--(0,2);
                \filldraw[fill=tensor] (-1,-1) rectangle (1,1);
                \draw (0,0) node {\scriptsize $A$};
            \end{tikzpicture}
        \end{array}
        =
        \begin{array}{c}
            \begin{tikzpicture}[scale=.2,thick]
                \draw (0,-3) node {};
                \draw (-5,0)--(5,0);
                \draw (0,0)--(0,5);        		  
                \filldraw[fill=tensor] (-1,-1) rectangle (1,1);
                \filldraw[fill=white] (-4,-1) rectangle (-2,1);
                \filldraw[fill=white] (2,-1) rectangle (4,1);
                \filldraw[fill=white] (-1,2) rectangle (1,4);
                \draw (0,0) node {\scriptsize $A$};
                \draw (-3,0) node {\scriptsize $X$};
                \draw (3,0) node {\scriptsize $X$};
                \draw (0,3) node {\scriptsize $X$};
            \end{tikzpicture}
        \end{array}.
        \label{eq:A_prop2}
    \end{align}
\end{subequations}
During the transformation of the MPS into its RG-based preparation circuit, these properties are inherited by the blocked tensors, the positive semidefinite tensors, and, in particular, the isometries.
The corresponding properties of  isometry $V$, given as
\begin{subequations}
    \begin{align}
        (i)&\quad
        \begin{array}{c}
            \begin{tikzpicture}[scale=.2,thick]
              \draw[white] (0,-5)--(0,5);
              \draw (1,-2)--(1,2);
              \draw (4,0)--(4,2);
              \draw (7,0)--(7,2);
              \draw (10,-2)--(10,2);
              \filldraw[fill=isometry] (0,-1) rectangle (11,1);
              \draw (5.5,0) node {\scriptsize $V$};
              \draw (1,3) node {\scriptsize $i$};
              \draw (1,-3) node {\scriptsize $j$};  
            \end{tikzpicture}
        \end{array}
        = 0 \qquad \forall \  i \neq j,\\
        (ii)&\quad
        \begin{array}{c}
            \begin{tikzpicture}[scale=.2,thick]
              \draw[white] (0,-5)--(0,5);
              \draw (1,-2)--(1,2);
              \draw (4,0)--(4,2);
              \draw (7,0)--(7,2);
              \draw (10,-2)--(10,2);
              \filldraw[fill=isometry] (0,-1) rectangle (11,1);
              \draw (5.5,0) node {\scriptsize $V$};
            \end{tikzpicture}
        \end{array}
        =
        \begin{array}{c}
            \begin{tikzpicture}[scale=.2,thick]
              \draw (1,-5)--(1,5);
              \draw (4,0)--(4,5);
              \draw (7,0)--(7,5);
              \draw (10,-5)--(10,5);
              \filldraw[fill=isometry] (0,-1) rectangle (11,1);
              \filldraw[fill=white] (0,2) rectangle (2,4);
              \filldraw[fill=white] (3,2) rectangle (5,4);
              \filldraw[fill=white] (6,2) rectangle (8,4);
              \filldraw[fill=white] (9,2) rectangle (11,4);
              \filldraw[fill=white] (0,-2) rectangle (2,-4);
              \filldraw[fill=white] (9,-2) rectangle (11,-4);
              \draw (5.5,0) node {\scriptsize $V$};
              \draw (1,3) node {\scriptsize $X$};
              \draw (4,3) node {\scriptsize $X$};
              \draw (7,3) node {\scriptsize $X$};
              \draw (10,3) node {\scriptsize $X$};
              \draw (1,-3) node {\scriptsize $X$};
              \draw (10,-3) node {\scriptsize $X$};
            \end{tikzpicture}
        \end{array}
    \end{align}
\end{subequations}
imply that $V$ written as a matrix in the computational basis is $(i)$ the direct sum of two $8 \times 2$ matrices and $(ii)$ point-symmetric with respect to its center. Thus, $V = \tilde{V}_1 \oplus \tilde{V}_2$ with $\tilde{V}_1, \tilde{V}_2 \in \mathbb{C}^{8 \times 2}$ and $(\tilde{V}_1)_{i,j} = (\tilde{V}_2)_{7-i,1-j} \ \forall \ i \in \{0,...,7\}, j \in \{0,1\}$. 

\begin{figure}[t]
  \centering
    \begin{quantikz}[scale=.2, thick]
        & \ctrl{3} & \qw & \ctrl{3} & \qw & \qw & \qw        \\
        \lstick{$\ket{0}$} & \qw & \gate[wires=3, style={fill=isometry,draw,minimum width=1cm}][1.2cm]{\tilde{V}_1} & \qw & \ctrl{2} & \qw & \qw        \\
        \lstick{$\ket{0}$} & \qw & \qw & \qw & \qw & \ctrl{1} & \qw  \\
        & \targ{}  & \qw  & \targ{} & \targ{} & \targ{} & \qw
    \end{quantikz}
    \caption{Quantum circuit implementing the $16 \times 4$ isometries of \cref{eq:approximated_mps}. With the implementation of $8 \times 2$ isometries proposed in Ref.~\cite{Iten_2016}, the overall CNOT count of 13 gates is a significant improvement over the 54 CNOT gates for an arbitrary $16 \times 4$ isometry~\cite{Iten_2016}.}
    \label{fig:isometry_implementation_using_block_structure}
\end{figure}
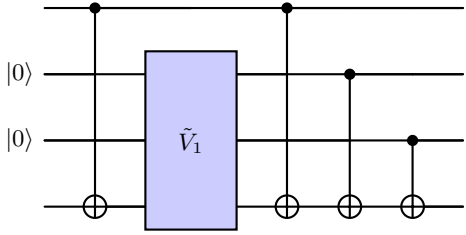

We exploit this structure in the implementation of isometry $V$ by placing 4 CNOT gates around the implementation of $\mathbb{I} \otimes \tilde{V}_1 = \tilde{V}_1 \oplus \tilde{V}_1$ to reverse the row and column order of the second summand, as shown in \cref{fig:isometry_implementation_using_block_structure}. As any $8 \times 2$ isometry can be implemented with 9 CNOT gates~\cite{Iten_2016}, our implementation consists of only 13 CNOT gates in total, corresponding to a significant improvement over the 54 CNOT gates for an arbitrary $16 \times 4$ isometry.

Additionally, this demonstrates that the depth of the RG-circuit is gauge-dependent. Gauges other than that of \cref{eq:A} generally do not satisfy the conditions of \cref{eq:A_prop1} and (\ref{eq:A_prop2}). In such cases, the corresponding isometries no longer inherit these properties and therefore do not allow for the more efficient implementation outlined above.

\subsection{Mapping}
\label{app:mapping}
\begin{figure}[t]
    \centering
    \includegraphics[width=.9\linewidth]{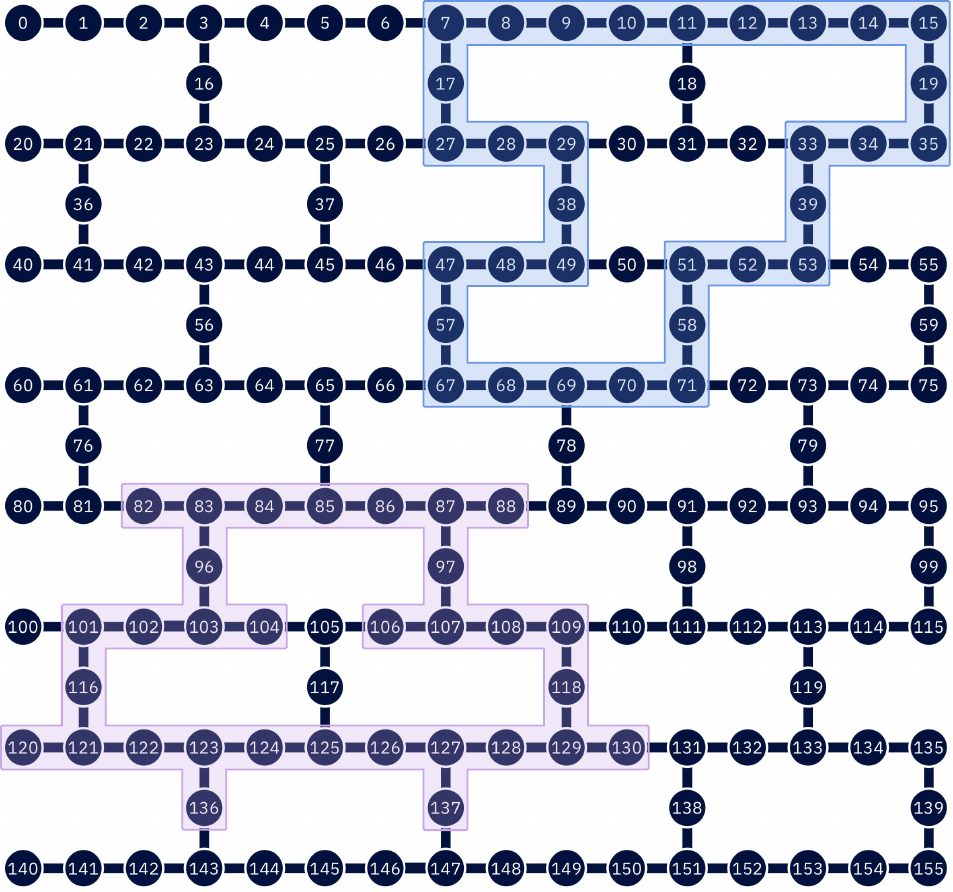}
    \caption{Layout of the \texttt{ibm\_fez} device. 156 superconducting qubits are arranged on a heavy-hex lattice, which is a hexagonal lattice with an additional node on each edge. Both the RG-based and sequential protocols are mapped to rings of qubits, as depicted in purple and blue, respectively. The ring used for the RG-based preparation consists of eight-qubit blocks, as illustrated in  \cref{fig:qubit_mapping}.}
    \label{fig:ibm_fez_layout}
\end{figure}

\begin{figure}[t]
    \centering
    \begin{equation*}
        \begin{array}{c}
            \begin{quantikz}[scale=.2, thick, row sep=0.2cm, column sep=0.5cm]
                &  \gate[wires=2]{\begin{smallmatrix} \text{PREP } \\ \ket{\omega} \end{smallmatrix}}  & \qw \\ 
                \lstick{$q_0$} & \ghost{\omega} &
                \gate[wires=8, style={fill=isometry, draw=black}]{V_8}  &
                \gate[wires=4, style={fill=isometry, draw=black}]{V_4} & \qw \\
                \lstick{$q_1$} & \qw & \ghost{V_8}  & \ghost{V_4} & \qw \\
                \lstick{$q_2$} & \qw & \ghost{V_8}  & \ghost{V_4} & \qw \\
                \lstick{$q_3$} & \qw & \ghost{V_8}  & \ghost{V_4} & \qw \\
                \lstick{$q_4$} & \qw & \ghost{V_8}  &
                \gate[wires=4, style={fill=isometry, draw=black}]{V_4} & \qw \\
                \lstick{$q_5$} & \qw & \ghost{V_8}  & \ghost{V_4} & \qw \\
                \lstick{$q_6$} & \qw & \ghost{V_8}  & \ghost{V_4} & \qw \\
                \lstick{$q_7$} & \gate[wires=2]{\begin{smallmatrix} \text{PREP } \\ \ket{\omega} \end{smallmatrix}} & \ghost{V_8}  & \ghost{V_4} & \qw \\
                & \ghost{\omega} & \qw
            \end{quantikz}%
        \end{array}
        \longrightarrow
        \begin{array}{c}
            \includegraphics[]{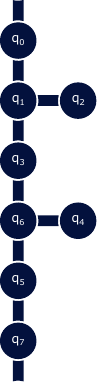}
        \end{array}
    \end{equation*}
    \caption{Mapping of identical eight-qubit blocks of the RG-based preparation circuit to \texttt{ibm\_fez}. This layout can be repeated to form a closed ring on the heavy-hex lattice for all $n \in \{16, 32, 48, 64, 80\}$.}
    \label{fig:qubit_mapping}
\end{figure}

A key step during the transpilation of a high-level quantum circuit to a specific quantum processor is the mapping of the virtual qubits of the circuit to the physical qubits of the target device. A good mapping minimizes the number of SWAP gates that need to be added to account for the limited connectivity of the quantum processor. On \texttt{ibm\_fez}, the superconducting qubits are arranged on a heavy-hex lattice (see \cref{fig:ibm_fez_layout}). We optimized the mapping of the RG-based  preparation circuit by partitioning it into identical blocks of eight neighboring qubits and mapping each of these blocks to a line with pendant vertices as shown in \cref{fig:qubit_mapping}. This line can be repeated on the heavy-hex lattice to form rings of sizes $n \in \{16, 32, 48, 64, 80\}$, as illustrated in \cref{fig:ibm_fez_layout}.

The mapping of the sequential preparation circuit to the target device was straightforward. It consists only of nearest-neighbor two-qubit gates. Thus, we transpiled the sequential circuits on simple rings, also illustrated in \cref{fig:ibm_fez_layout}.

\begin{figure}[htb]
  \centering
    \includegraphics[width=\linewidth]{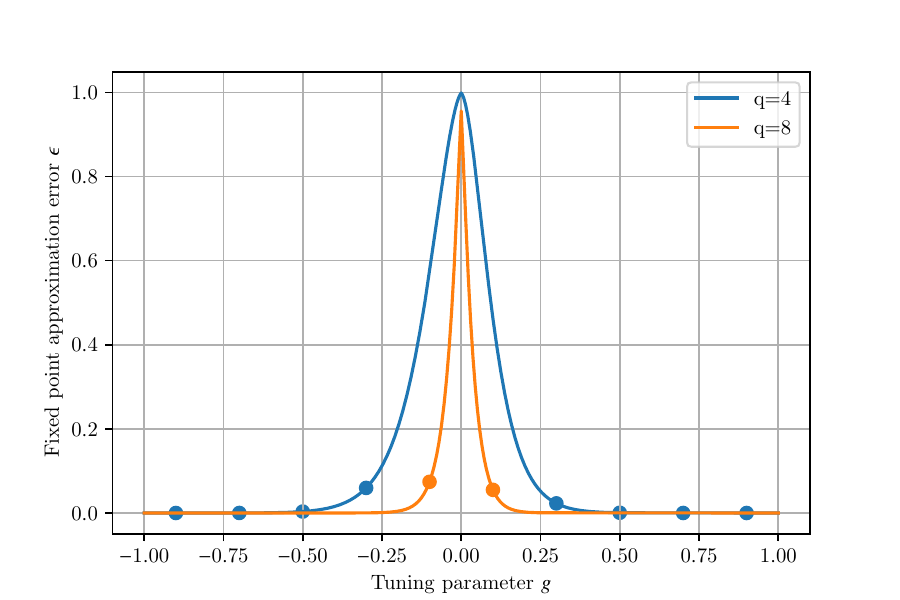}
    \caption{Fixed point approximation error for blocking numbers $q=4$ and $q=8$ for the MPS of \cref{eq:A} with $n=80$ sites. As the correlation length increases for $|g| \to 0$, so does the error resulting from the fixed point approximation in the RG-based protocol. Therefore, we increased the blocking number from $q=4$ to $q=8$ for the two points closest to the phase transition ($g=\pm0.1$), resulting in approximation errors for the states we prepare here, indicated by the colored dots, of at most $7.5\%$.}
    \label{fig:fixed_point_approximation_error}
\end{figure}

\bibliography{biblio,library}

\begin{thebibliography}{30}%
\makeatletter
\providecommand \@ifxundefined [1]{%
 \@ifx{#1\undefined}
}%
\providecommand \@ifnum [1]{%
 \ifnum #1\expandafter \@firstoftwo
 \else \expandafter \@secondoftwo
 \fi
}%
\providecommand \@ifx [1]{%
 \ifx #1\expandafter \@firstoftwo
 \else \expandafter \@secondoftwo
 \fi
}%
\providecommand \natexlab [1]{#1}%
\providecommand \enquote  [1]{``#1''}%
\providecommand \bibnamefont  [1]{#1}%
\providecommand \bibfnamefont [1]{#1}%
\providecommand \citenamefont [1]{#1}%
\providecommand \href@noop [0]{\@secondoftwo}%
\providecommand \href [0]{\begingroup \@sanitize@url \@href}%
\providecommand \@href[1]{\@@startlink{#1}\@@href}%
\providecommand \@@href[1]{\endgroup#1\@@endlink}%
\providecommand \@sanitize@url [0]{\catcode `\\12\catcode `\$12\catcode `\&12\catcode `\#12\catcode `\^12\catcode `\_12\catcode `\%12\relax}%
\providecommand \@@startlink[1]{}%
\providecommand \@@endlink[0]{}%
\providecommand \url  [0]{\begingroup\@sanitize@url \@url }%
\providecommand \@url [1]{\endgroup\@href {#1}{\urlprefix }}%
\providecommand \urlprefix  [0]{URL }%
\providecommand \Eprint [0]{\href }%
\providecommand \doibase [0]{https://doi.org/}%
\providecommand \selectlanguage [0]{\@gobble}%
\providecommand \bibinfo  [0]{\@secondoftwo}%
\providecommand \bibfield  [0]{\@secondoftwo}%
\providecommand \translation [1]{[#1]}%
\providecommand \BibitemOpen [0]{}%
\providecommand \bibitemStop [0]{}%
\providecommand \bibitemNoStop [0]{.\EOS\space}%
\providecommand \EOS [0]{\spacefactor3000\relax}%
\providecommand \BibitemShut  [1]{\csname bibitem#1\endcsname}%
\let\auto@bib@innerbib\@empty
\bibitem [{\citenamefont {Malz}\ \emph {et~al.}(2024)\citenamefont {Malz}, \citenamefont {Styliaris}, \citenamefont {Wei},\ and\ \citenamefont {Cirac}}]{Malz_2024}%
  \BibitemOpen
  \bibfield  {author} {\bibinfo {author} {\bibfnamefont {D.}~\bibnamefont {Malz}}, \bibinfo {author} {\bibfnamefont {G.}~\bibnamefont {Styliaris}}, \bibinfo {author} {\bibfnamefont {Z.-Y.}\ \bibnamefont {Wei}},\ and\ \bibinfo {author} {\bibfnamefont {J.~I.}\ \bibnamefont {Cirac}},\ }\bibfield  {title} {\bibinfo {title} {Preparation of matrix product states with log-depth quantum circuits},\ }\href {https://doi.org/10.1103/PhysRevLett.132.040404} {\bibfield  {journal} {\bibinfo  {journal} {Phys. Rev. Lett.}\ }\textbf {\bibinfo {volume} {132}},\ \bibinfo {pages} {040404} (\bibinfo {year} {2024})}\BibitemShut {NoStop}%
\bibitem [{\citenamefont {{Fannes}}\ \emph {et~al.}(1992)\citenamefont {{Fannes}}, \citenamefont {{Nachtergaele}},\ and\ \citenamefont {{Werner}}}]{Fannes_1992}%
  \BibitemOpen
  \bibfield  {author} {\bibinfo {author} {\bibfnamefont {M.}~\bibnamefont {{Fannes}}}, \bibinfo {author} {\bibfnamefont {B.}~\bibnamefont {{Nachtergaele}}},\ and\ \bibinfo {author} {\bibfnamefont {R.~F.}\ \bibnamefont {{Werner}}},\ }\bibfield  {title} {\bibinfo {title} {{Finitely correlated states on quantum spin chains}},\ }\href {https://doi.org/10.1007/BF02099178} {\bibfield  {journal} {\bibinfo  {journal} {Communications in Mathematical Physics}\ }\textbf {\bibinfo {volume} {144}},\ \bibinfo {pages} {443} (\bibinfo {year} {1992})}\BibitemShut {NoStop}%
\bibitem [{\citenamefont {Hastings}(2007)}]{Hastings_2007}%
  \BibitemOpen
  \bibfield  {author} {\bibinfo {author} {\bibfnamefont {M.~B.}\ \bibnamefont {Hastings}},\ }\bibfield  {title} {\bibinfo {title} {{An area law for one-dimensional quantum systems}},\ }\href {https://doi.org/10.1088/1742-5468/2007/08/P08024} {\bibfield  {journal} {\bibinfo  {journal} {J. Stat. Mech.: Theory Exp.}\ }\textbf {\bibinfo {volume} {2007}}\bibinfo  {number} { (08)},\ \bibinfo {pages} {P08024}}\BibitemShut {NoStop}%
\bibitem [{\citenamefont {McCulloch}(2007)}]{McCulloch_2007}%
  \BibitemOpen
\bibfield  {number} {  }\bibfield  {author} {\bibinfo {author} {\bibfnamefont {I.~P.}\ \bibnamefont {McCulloch}},\ }\bibfield  {title} {\bibinfo {title} {From density-matrix renormalization group to matrix product states},\ }\href {https://doi.org/10.1088/1742-5468/2007/10/p10014} {\bibfield  {journal} {\bibinfo  {journal} {J. Stat. Mech.: Theory Exp.}\ }\textbf {\bibinfo {volume} {2007}}\bibinfo  {number} { (10)},\ \bibinfo {pages} {P10014–P10014}}\BibitemShut {NoStop}%
\bibitem [{\citenamefont {Greenberger}\ \emph {et~al.}(1989)\citenamefont {Greenberger}, \citenamefont {Horne},\ and\ \citenamefont {Zeilinger}}]{Greenberger1989}%
  \BibitemOpen
\bibfield  {number} {  }\bibfield  {author} {\bibinfo {author} {\bibfnamefont {D.~M.}\ \bibnamefont {Greenberger}}, \bibinfo {author} {\bibfnamefont {M.~A.}\ \bibnamefont {Horne}},\ and\ \bibinfo {author} {\bibfnamefont {A.}~\bibnamefont {Zeilinger}},\ }\bibinfo {title} {Going beyond bell's theorem},\ in\ \href {https://doi.org/10.1007/978-94-017-0849-4_10} {\emph {\bibinfo {booktitle} {Bell's Theorem, Quantum Theory and Conceptions of the Universe}}}\ (\bibinfo  {publisher} {Springer Netherlands},\ \bibinfo {address} {Dordrecht},\ \bibinfo {year} {1989})\ pp.\ \bibinfo {pages} {69--72}\BibitemShut {NoStop}%
\bibitem [{\citenamefont {D\"ur}\ \emph {et~al.}(2000)\citenamefont {D\"ur}, \citenamefont {Vidal},\ and\ \citenamefont {Cirac}}]{Dur_2000}%
  \BibitemOpen
  \bibfield  {author} {\bibinfo {author} {\bibfnamefont {W.}~\bibnamefont {D\"ur}}, \bibinfo {author} {\bibfnamefont {G.}~\bibnamefont {Vidal}},\ and\ \bibinfo {author} {\bibfnamefont {J.~I.}\ \bibnamefont {Cirac}},\ }\bibfield  {title} {\bibinfo {title} {Three qubits can be entangled in two inequivalent ways},\ }\href {https://doi.org/10.1103/PhysRevA.62.062314} {\bibfield  {journal} {\bibinfo  {journal} {Phys. Rev. A}\ }\textbf {\bibinfo {volume} {62}},\ \bibinfo {pages} {062314} (\bibinfo {year} {2000})}\BibitemShut {NoStop}%
\bibitem [{\citenamefont {Briegel}\ and\ \citenamefont {Raussendorf}(2001)}]{Briegel_2001}%
  \BibitemOpen
  \bibfield  {author} {\bibinfo {author} {\bibfnamefont {H.~J.}\ \bibnamefont {Briegel}}\ and\ \bibinfo {author} {\bibfnamefont {R.}~\bibnamefont {Raussendorf}},\ }\bibfield  {title} {\bibinfo {title} {Persistent entanglement in arrays of interacting particles},\ }\href {https://doi.org/10.1103/PhysRevLett.86.910} {\bibfield  {journal} {\bibinfo  {journal} {Phys. Rev. Lett.}\ }\textbf {\bibinfo {volume} {86}},\ \bibinfo {pages} {910} (\bibinfo {year} {2001})}\BibitemShut {NoStop}%
\bibitem [{\citenamefont {Affleck}\ \emph {et~al.}(1987)\citenamefont {Affleck}, \citenamefont {Kennedy}, \citenamefont {Lieb},\ and\ \citenamefont {Tasaki}}]{Affleck_1987}%
  \BibitemOpen
  \bibfield  {author} {\bibinfo {author} {\bibfnamefont {I.}~\bibnamefont {Affleck}}, \bibinfo {author} {\bibfnamefont {T.}~\bibnamefont {Kennedy}}, \bibinfo {author} {\bibfnamefont {E.~H.}\ \bibnamefont {Lieb}},\ and\ \bibinfo {author} {\bibfnamefont {H.}~\bibnamefont {Tasaki}},\ }\bibfield  {title} {\bibinfo {title} {Rigorous results on valence-bond ground states in antiferromagnets},\ }\href {https://doi.org/10.1103/PhysRevLett.59.799} {\bibfield  {journal} {\bibinfo  {journal} {Phys. Rev. Lett.}\ }\textbf {\bibinfo {volume} {59}},\ \bibinfo {pages} {799} (\bibinfo {year} {1987})}\BibitemShut {NoStop}%
\bibitem [{\citenamefont {Ran}(2020)}]{Ran2020_EncodingMPS}%
  \BibitemOpen
  \bibfield  {author} {\bibinfo {author} {\bibfnamefont {S.-J.}\ \bibnamefont {Ran}},\ }\bibfield  {title} {\bibinfo {title} {{Encoding of matrix product states into quantum circuits of one- and two-qubit gates}},\ }\href {https://doi.org/10.1103/PhysRevA.101.032310} {\bibfield  {journal} {\bibinfo  {journal} {Phys. Rev. A}\ }\textbf {\bibinfo {volume} {101}},\ \bibinfo {pages} {032310} (\bibinfo {year} {2020})}\BibitemShut {NoStop}%
\bibitem [{\citenamefont {Rudolph}\ \emph {et~al.}(2023)\citenamefont {Rudolph}, \citenamefont {Chen}, \citenamefont {Miller}, \citenamefont {Acharya},\ and\ \citenamefont {Perdomo-Ortiz}}]{Rudolph2023_MPS}%
  \BibitemOpen
  \bibfield  {author} {\bibinfo {author} {\bibfnamefont {M.~S.}\ \bibnamefont {Rudolph}}, \bibinfo {author} {\bibfnamefont {J.}~\bibnamefont {Chen}}, \bibinfo {author} {\bibfnamefont {J.}~\bibnamefont {Miller}}, \bibinfo {author} {\bibfnamefont {A.}~\bibnamefont {Acharya}},\ and\ \bibinfo {author} {\bibfnamefont {A.}~\bibnamefont {Perdomo-Ortiz}},\ }\bibfield  {title} {\bibinfo {title} {{Decomposition of matrix product states into shallow quantum circuits}},\ }\href {https://doi.org/10.1088/2058-9565/ad04e6} {\bibfield  {journal} {\bibinfo  {journal} {Quantum Sci. Technol.}\ }\textbf {\bibinfo {volume} {9}},\ \bibinfo {pages} {015012} (\bibinfo {year} {2023})}\BibitemShut {NoStop}%
\bibitem [{\citenamefont {Iqbal}\ \emph {et~al.}(2022)\citenamefont {Iqbal}, \citenamefont {Ramo},\ and\ \citenamefont {Dreyer}}]{iqbal2022preentangling}%
  \BibitemOpen
  \bibfield  {author} {\bibinfo {author} {\bibfnamefont {M.}~\bibnamefont {Iqbal}}, \bibinfo {author} {\bibfnamefont {D.~M.}\ \bibnamefont {Ramo}},\ and\ \bibinfo {author} {\bibfnamefont {H.}~\bibnamefont {Dreyer}},\ }\bibfield  {title} {\bibinfo {title} {Preentangling quantum algorithms--the density matrix renormalization group-assisted quantum canonical transformation},\ }\href {https://arxiv.org/abs/2209.07106} {\bibfield  {journal} {\bibinfo  {journal} {arXiv:2209.07106}\ } (\bibinfo {year} {2022})}\BibitemShut {NoStop}%
\bibitem [{\citenamefont {Wei}\ and\ \citenamefont {Malz}(2025)}]{wei2025state2}%
  \BibitemOpen
  \bibfield  {author} {\bibinfo {author} {\bibfnamefont {Z.-Y.}\ \bibnamefont {Wei}}\ and\ \bibinfo {author} {\bibfnamefont {D.}~\bibnamefont {Malz}},\ }\bibfield  {title} {\bibinfo {title} {State preparation with parallel-sequential circuits},\ }\href {https://arxiv.org/abs/2503.14645} {\bibfield  {journal} {\bibinfo  {journal} {arXiv preprint arXiv:2503.14645}\ } (\bibinfo {year} {2025})}\BibitemShut {NoStop}%
\bibitem [{\citenamefont {Sch\"on}\ \emph {et~al.}(2005)\citenamefont {Sch\"on}, \citenamefont {Solano}, \citenamefont {Verstraete}, \citenamefont {Cirac},\ and\ \citenamefont {Wolf}}]{Schoen_2005}%
  \BibitemOpen
  \bibfield  {author} {\bibinfo {author} {\bibfnamefont {C.}~\bibnamefont {Sch\"on}}, \bibinfo {author} {\bibfnamefont {E.}~\bibnamefont {Solano}}, \bibinfo {author} {\bibfnamefont {F.}~\bibnamefont {Verstraete}}, \bibinfo {author} {\bibfnamefont {J.~I.}\ \bibnamefont {Cirac}},\ and\ \bibinfo {author} {\bibfnamefont {M.~M.}\ \bibnamefont {Wolf}},\ }\bibfield  {title} {\bibinfo {title} {Sequential generation of entangled multiqubit states},\ }\href {https://doi.org/10.1103/PhysRevLett.95.110503} {\bibfield  {journal} {\bibinfo  {journal} {Phys. Rev. Lett.}\ }\textbf {\bibinfo {volume} {95}},\ \bibinfo {pages} {110503} (\bibinfo {year} {2005})}\BibitemShut {NoStop}%
\bibitem [{\citenamefont {Smith}\ \emph {et~al.}(2022)\citenamefont {Smith}, \citenamefont {Jobst}, \citenamefont {Green},\ and\ \citenamefont {Pollmann}}]{smith2022crossing}%
  \BibitemOpen
  \bibfield  {author} {\bibinfo {author} {\bibfnamefont {A.}~\bibnamefont {Smith}}, \bibinfo {author} {\bibfnamefont {B.}~\bibnamefont {Jobst}}, \bibinfo {author} {\bibfnamefont {A.~G.}\ \bibnamefont {Green}},\ and\ \bibinfo {author} {\bibfnamefont {F.}~\bibnamefont {Pollmann}},\ }\bibfield  {title} {\bibinfo {title} {{Crossing a topological phase transition with a quantum computer}},\ }\href {https://journals.aps.org/prresearch/abstract/10.1103/PhysRevResearch.4.L022020} {\bibfield  {journal} {\bibinfo  {journal} {Physical Review Research}\ }\textbf {\bibinfo {volume} {4}},\ \bibinfo {pages} {L022020} (\bibinfo {year} {2022})}\BibitemShut {NoStop}%
\bibitem [{\citenamefont {Zhang}\ \emph {et~al.}(2022)\citenamefont {Zhang}, \citenamefont {Jahanbani}, \citenamefont {Niu}, \citenamefont {Haghshenas},\ and\ \citenamefont {Potter}}]{Potter2022_qMPO}%
  \BibitemOpen
  \bibfield  {author} {\bibinfo {author} {\bibfnamefont {Y.}~\bibnamefont {Zhang}}, \bibinfo {author} {\bibfnamefont {S.}~\bibnamefont {Jahanbani}}, \bibinfo {author} {\bibfnamefont {D.}~\bibnamefont {Niu}}, \bibinfo {author} {\bibfnamefont {R.}~\bibnamefont {Haghshenas}},\ and\ \bibinfo {author} {\bibfnamefont {A.~C.}\ \bibnamefont {Potter}},\ }\bibfield  {title} {\bibinfo {title} {{Qubit-efficient simulation of thermal states with quantum tensor networks}},\ }\href {https://doi.org/10.1103/PhysRevB.106.165126} {\bibfield  {journal} {\bibinfo  {journal} {Phys. Rev. B}\ }\textbf {\bibinfo {volume} {106}},\ \bibinfo {pages} {165126} (\bibinfo {year} {2022})}\BibitemShut {NoStop}%
\bibitem [{\citenamefont {Wall}\ \emph {et~al.}(2024)\citenamefont {Wall}, \citenamefont {Reilly}, \citenamefont {Van~Dyke}, \citenamefont {Broholm},\ and\ \citenamefont {Titum}}]{Wall2024_qMPS-Dynamics}%
  \BibitemOpen
  \bibfield  {author} {\bibinfo {author} {\bibfnamefont {M.~L.}\ \bibnamefont {Wall}}, \bibinfo {author} {\bibfnamefont {A.}~\bibnamefont {Reilly}}, \bibinfo {author} {\bibfnamefont {J.~S.}\ \bibnamefont {Van~Dyke}}, \bibinfo {author} {\bibfnamefont {C.}~\bibnamefont {Broholm}},\ and\ \bibinfo {author} {\bibfnamefont {P.}~\bibnamefont {Titum}},\ }\bibfield  {title} {\bibinfo {title} {Quantum tensor network algorithms for evaluation of spectral functions on quantum computers},\ }\href {https://doi.org/10.1103/PhysRevB.110.214402} {\bibfield  {journal} {\bibinfo  {journal} {Phys. Rev. B}\ }\textbf {\bibinfo {volume} {110}},\ \bibinfo {pages} {214402} (\bibinfo {year} {2024})}\BibitemShut {NoStop}%
\bibitem [{\citenamefont {Anselme~Martin}\ \emph {et~al.}(2024)\citenamefont {Anselme~Martin}, \citenamefont {Ayral}, \citenamefont {Jamet}, \citenamefont {Ran\v{c}i\'{c}},\ and\ \citenamefont {Simon}}]{Anselme2024_MPS-Trotter}%
  \BibitemOpen
  \bibfield  {author} {\bibinfo {author} {\bibfnamefont {B.}~\bibnamefont {Anselme~Martin}}, \bibinfo {author} {\bibfnamefont {T.}~\bibnamefont {Ayral}}, \bibinfo {author} {\bibfnamefont {F.}~\bibnamefont {Jamet}}, \bibinfo {author} {\bibfnamefont {M.~J.}\ \bibnamefont {Ran\v{c}i\'{c}}},\ and\ \bibinfo {author} {\bibfnamefont {P.}~\bibnamefont {Simon}},\ }\bibfield  {title} {\bibinfo {title} {{Combining matrix product states and noisy quantum computers for quantum simulation}},\ }\href {https://doi.org/10.1103/PhysRevA.109.062437} {\bibfield  {journal} {\bibinfo  {journal} {Phys. Rev. A}\ }\textbf {\bibinfo {volume} {109}},\ \bibinfo {pages} {062437} (\bibinfo {year} {2024})}\BibitemShut {NoStop}%
\bibitem [{\citenamefont {Schuhmacher}\ \emph {et~al.}(2025)\citenamefont {Schuhmacher}, \citenamefont {Ballarin}, \citenamefont {Baiardi}, \citenamefont {Magnifico}, \citenamefont {Tacchino}, \citenamefont {Montangero},\ and\ \citenamefont {Tavernelli}}]{Schuhmacher2025_hTTN}%
  \BibitemOpen
  \bibfield  {author} {\bibinfo {author} {\bibfnamefont {J.}~\bibnamefont {Schuhmacher}}, \bibinfo {author} {\bibfnamefont {M.}~\bibnamefont {Ballarin}}, \bibinfo {author} {\bibfnamefont {A.}~\bibnamefont {Baiardi}}, \bibinfo {author} {\bibfnamefont {G.}~\bibnamefont {Magnifico}}, \bibinfo {author} {\bibfnamefont {F.}~\bibnamefont {Tacchino}}, \bibinfo {author} {\bibfnamefont {S.}~\bibnamefont {Montangero}},\ and\ \bibinfo {author} {\bibfnamefont {I.}~\bibnamefont {Tavernelli}},\ }\bibfield  {title} {\bibinfo {title} {{Hybrid Tree Tensor Networks for Quantum Simulation}},\ }\href {https://doi.org/10.1103/PRXQuantum.6.010320} {\bibfield  {journal} {\bibinfo  {journal} {PRX Quantum}\ }\textbf {\bibinfo {volume} {6}},\ \bibinfo {pages} {010320} (\bibinfo {year} {2025})}\BibitemShut {NoStop}%
\bibitem [{\citenamefont {Smith}\ \emph {et~al.}(2023)\citenamefont {Smith}, \citenamefont {Crane}, \citenamefont {Wiebe},\ and\ \citenamefont {Girvin}}]{Smith2023_AKLT-Fusion}%
  \BibitemOpen
  \bibfield  {author} {\bibinfo {author} {\bibfnamefont {K.~C.}\ \bibnamefont {Smith}}, \bibinfo {author} {\bibfnamefont {E.}~\bibnamefont {Crane}}, \bibinfo {author} {\bibfnamefont {N.}~\bibnamefont {Wiebe}},\ and\ \bibinfo {author} {\bibfnamefont {S.}~\bibnamefont {Girvin}},\ }\bibfield  {title} {\bibinfo {title} {{Deterministic Constant-Depth Preparation of the AKLT State on a Quantum Processor Using Fusion Measurements}},\ }\bibfield  {journal} {\bibinfo  {journal} {PRX Quantum}\ }\textbf {\bibinfo {volume} {4}},\ \href {https://doi.org/10.1103/prxquantum.4.020315} {10.1103/prxquantum.4.020315} (\bibinfo {year} {2023})\BibitemShut {NoStop}%
\bibitem [{\citenamefont {Sahay}\ and\ \citenamefont {Verresen}(2024)}]{Sahay2024_FiniteDepth-TNS}%
  \BibitemOpen
  \bibfield  {author} {\bibinfo {author} {\bibfnamefont {R.}~\bibnamefont {Sahay}}\ and\ \bibinfo {author} {\bibfnamefont {R.}~\bibnamefont {Verresen}},\ }\bibfield  {title} {\bibinfo {title} {{Finite-Depth Preparation of Tensor Network States from Measurement}},\ }\href {https://arxiv.org/abs/2404.17087} {\bibfield  {journal} {\bibinfo  {journal} {arXiv}\ ,\ \bibinfo {pages} {2404.17087}} (\bibinfo {year} {2024})}\BibitemShut {NoStop}%
\bibitem [{\citenamefont {Stephen}\ and\ \citenamefont {Hart}(2024)}]{Stephen2024_MPS-Fusion-Characterization}%
  \BibitemOpen
  \bibfield  {author} {\bibinfo {author} {\bibfnamefont {D.~T.}\ \bibnamefont {Stephen}}\ and\ \bibinfo {author} {\bibfnamefont {O.}~\bibnamefont {Hart}},\ }\bibfield  {title} {\bibinfo {title} {{Preparing matrix product states via fusion: constraints and extensions}},\ }\href {https://arxiv.org/abs/2404.16360} {\bibfield  {journal} {\bibinfo  {journal} {arXiv}\ ,\ \bibinfo {pages} {2404.16360}} (\bibinfo {year} {2024})}\BibitemShut {NoStop}%
\bibitem [{\citenamefont {Zhang}\ \emph {et~al.}(2024)\citenamefont {Zhang}, \citenamefont {Gopalakrishnan},\ and\ \citenamefont {Styliaris}}]{Styliaris2024_Fusion-PEPS}%
  \BibitemOpen
  \bibfield  {author} {\bibinfo {author} {\bibfnamefont {Y.}~\bibnamefont {Zhang}}, \bibinfo {author} {\bibfnamefont {S.}~\bibnamefont {Gopalakrishnan}},\ and\ \bibinfo {author} {\bibfnamefont {G.}~\bibnamefont {Styliaris}},\ }\bibfield  {title} {\bibinfo {title} {{Characterizing Matrix-Product States and Projected Entangled-Pair States Preparable via Measurement and Feedback}},\ }\href {https://doi.org/10.1103/PRXQuantum.5.040304} {\bibfield  {journal} {\bibinfo  {journal} {PRX Quantum}\ }\textbf {\bibinfo {volume} {5}},\ \bibinfo {pages} {040304} (\bibinfo {year} {2024})}\BibitemShut {NoStop}%
\bibitem [{\citenamefont {Ge}\ \emph {et~al.}(2016)\citenamefont {Ge}, \citenamefont {Moln{\'{a}}r},\ and\ \citenamefont {Cirac}}]{Ge2016}%
  \BibitemOpen
  \bibfield  {author} {\bibinfo {author} {\bibfnamefont {Y.}~\bibnamefont {Ge}}, \bibinfo {author} {\bibfnamefont {A.}~\bibnamefont {Moln{\'{a}}r}},\ and\ \bibinfo {author} {\bibfnamefont {J.~I.}\ \bibnamefont {Cirac}},\ }\bibfield  {title} {\bibinfo {title} {{Rapid Adiabatic Preparation of Injective Projected Entangled Pair States and Gibbs States}},\ }\href {https://doi.org/10.1103/PhysRevLett.116.080503} {\bibfield  {journal} {\bibinfo  {journal} {Physical Review Letters}\ }\textbf {\bibinfo {volume} {116}},\ \bibinfo {pages} {1} (\bibinfo {year} {2016})}\BibitemShut {NoStop}%
\bibitem [{\citenamefont {Wei}\ \emph {et~al.}(2023)\citenamefont {Wei}, \citenamefont {Malz},\ and\ \citenamefont {Cirac}}]{zyadi}%
  \BibitemOpen
  \bibfield  {author} {\bibinfo {author} {\bibfnamefont {Z.-Y.}\ \bibnamefont {Wei}}, \bibinfo {author} {\bibfnamefont {D.}~\bibnamefont {Malz}},\ and\ \bibinfo {author} {\bibfnamefont {J.~I.}\ \bibnamefont {Cirac}},\ }\bibfield  {title} {\bibinfo {title} {Efficient adiabatic preparation of tensor network states},\ }\href {https://journals.aps.org/prresearch/abstract/10.1103/PhysRevResearch.5.L022037} {\bibfield  {journal} {\bibinfo  {journal} {Physical Review Research}\ }\textbf {\bibinfo {volume} {5}},\ \bibinfo {pages} {L022037} (\bibinfo {year} {2023})}\BibitemShut {NoStop}%
\bibitem [{\citenamefont {Wolf}\ \emph {et~al.}(2006)\citenamefont {Wolf}, \citenamefont {Ortiz}, \citenamefont {Verstraete},\ and\ \citenamefont {Cirac}}]{Wolf_2006}%
  \BibitemOpen
  \bibfield  {author} {\bibinfo {author} {\bibfnamefont {M.~M.}\ \bibnamefont {Wolf}}, \bibinfo {author} {\bibfnamefont {G.}~\bibnamefont {Ortiz}}, \bibinfo {author} {\bibfnamefont {F.}~\bibnamefont {Verstraete}},\ and\ \bibinfo {author} {\bibfnamefont {J.~I.}\ \bibnamefont {Cirac}},\ }\bibfield  {title} {\bibinfo {title} {Quantum phase transitions in matrix product systems},\ }\href {https://doi.org/10.1103/PhysRevLett.97.110403} {\bibfield  {journal} {\bibinfo  {journal} {Phys. Rev. Lett.}\ }\textbf {\bibinfo {volume} {97}},\ \bibinfo {pages} {110403} (\bibinfo {year} {2006})}\BibitemShut {NoStop}%
\bibitem [{\citenamefont {Smith}\ \emph {et~al.}(2024)\citenamefont {Smith}, \citenamefont {Khan}, \citenamefont {Clark}, \citenamefont {Girvin},\ and\ \citenamefont {Wei}}]{Smith2024mps}%
  \BibitemOpen
  \bibfield  {author} {\bibinfo {author} {\bibfnamefont {K.~C.}\ \bibnamefont {Smith}}, \bibinfo {author} {\bibfnamefont {A.}~\bibnamefont {Khan}}, \bibinfo {author} {\bibfnamefont {B.~K.}\ \bibnamefont {Clark}}, \bibinfo {author} {\bibfnamefont {S.}~\bibnamefont {Girvin}},\ and\ \bibinfo {author} {\bibfnamefont {T.-C.}\ \bibnamefont {Wei}},\ }\bibfield  {title} {\bibinfo {title} {Constant-depth preparation of matrix product states with adaptive quantum circuits},\ }\href {https://doi.org/10.1103/PRXQuantum.5.030344} {\bibfield  {journal} {\bibinfo  {journal} {PRX Quantum}\ }\textbf {\bibinfo {volume} {5}},\ \bibinfo {pages} {030344} (\bibinfo {year} {2024})}\BibitemShut {NoStop}%
\bibitem [{\citenamefont {Verstraete}\ \emph {et~al.}(2005)\citenamefont {Verstraete}, \citenamefont {Cirac}, \citenamefont {Latorre}, \citenamefont {Rico},\ and\ \citenamefont {Wolf}}]{Verstraete2005}%
  \BibitemOpen
  \bibfield  {author} {\bibinfo {author} {\bibfnamefont {F.}~\bibnamefont {Verstraete}}, \bibinfo {author} {\bibfnamefont {J.~I.}\ \bibnamefont {Cirac}}, \bibinfo {author} {\bibfnamefont {J.~I.}\ \bibnamefont {Latorre}}, \bibinfo {author} {\bibfnamefont {E.}~\bibnamefont {Rico}},\ and\ \bibinfo {author} {\bibfnamefont {M.~M.}\ \bibnamefont {Wolf}},\ }\bibfield  {title} {\bibinfo {title} {{Renormalization-group transformations on quantum states}},\ }\href {https://doi.org/10.1103/PhysRevLett.94.140601} {\bibfield  {journal} {\bibinfo  {journal} {Physical Review Letters}\ }\textbf {\bibinfo {volume} {94}},\ \bibinfo {pages} {1} (\bibinfo {year} {2005})}\BibitemShut {NoStop}%
\bibitem [{\citenamefont {Iten}\ \emph {et~al.}(2016)\citenamefont {Iten}, \citenamefont {Colbeck}, \citenamefont {Kukuljan}, \citenamefont {Home},\ and\ \citenamefont {Christandl}}]{Iten_2016}%
  \BibitemOpen
  \bibfield  {author} {\bibinfo {author} {\bibfnamefont {R.}~\bibnamefont {Iten}}, \bibinfo {author} {\bibfnamefont {R.}~\bibnamefont {Colbeck}}, \bibinfo {author} {\bibfnamefont {I.}~\bibnamefont {Kukuljan}}, \bibinfo {author} {\bibfnamefont {J.}~\bibnamefont {Home}},\ and\ \bibinfo {author} {\bibfnamefont {M.}~\bibnamefont {Christandl}},\ }\bibfield  {title} {\bibinfo {title} {Quantum circuits for isometries},\ }\href {https://doi.org/10.1103/PhysRevA.93.032318} {\bibfield  {journal} {\bibinfo  {journal} {Phys. Rev. A}\ }\textbf {\bibinfo {volume} {93}},\ \bibinfo {pages} {032318} (\bibinfo {year} {2016})}\BibitemShut {NoStop}%
\bibitem [{\citenamefont {Javadi-Abhari}\ \emph {et~al.}(2024)\citenamefont {Javadi-Abhari}, \citenamefont {Treinish}, \citenamefont {Krsulich}, \citenamefont {Wood}, \citenamefont {Lishman}, \citenamefont {Gacon}, \citenamefont {Martiel}, \citenamefont {Nation}, \citenamefont {Bishop}, \citenamefont {Cross}, \citenamefont {Johnson},\ and\ \citenamefont {Gambetta}}]{qiskit2024}%
  \BibitemOpen
  \bibfield  {author} {\bibinfo {author} {\bibfnamefont {A.}~\bibnamefont {Javadi-Abhari}}, \bibinfo {author} {\bibfnamefont {M.}~\bibnamefont {Treinish}}, \bibinfo {author} {\bibfnamefont {K.}~\bibnamefont {Krsulich}}, \bibinfo {author} {\bibfnamefont {C.~J.}\ \bibnamefont {Wood}}, \bibinfo {author} {\bibfnamefont {J.}~\bibnamefont {Lishman}}, \bibinfo {author} {\bibfnamefont {J.}~\bibnamefont {Gacon}}, \bibinfo {author} {\bibfnamefont {S.}~\bibnamefont {Martiel}}, \bibinfo {author} {\bibfnamefont {P.~D.}\ \bibnamefont {Nation}}, \bibinfo {author} {\bibfnamefont {L.~S.}\ \bibnamefont {Bishop}}, \bibinfo {author} {\bibfnamefont {A.~W.}\ \bibnamefont {Cross}}, \bibinfo {author} {\bibfnamefont {B.~R.}\ \bibnamefont {Johnson}},\ and\ \bibinfo {author} {\bibfnamefont {J.~M.}\ \bibnamefont {Gambetta}},\ }\href {https://doi.org/10.48550/arXiv.2405.08810} {\bibinfo {title} {Quantum computing with {Q}iskit}} (\bibinfo {year} {2024}),\ \Eprint {https://arxiv.org/abs/2405.08810} {arXiv:2405.08810 [quant-ph]}
  \BibitemShut {NoStop}%
\bibitem [{\citenamefont {Jaderberg}\ \emph {et~al.}(2025)\citenamefont {Jaderberg}, \citenamefont {Pennington}, \citenamefont {Marshall}, \citenamefont {Anderson}, \citenamefont {Agarwal}, \citenamefont {Lindoy}, \citenamefont {Rungger}, \citenamefont {Mensa},\ and\ \citenamefont {Crain}}]{Jaderberg2025_VariationalPreparationNormalMPS}%
  \BibitemOpen
  \bibfield  {author} {\bibinfo {author} {\bibfnamefont {B.}~\bibnamefont {Jaderberg}}, \bibinfo {author} {\bibfnamefont {G.}~\bibnamefont {Pennington}}, \bibinfo {author} {\bibfnamefont {K.~V.}\ \bibnamefont {Marshall}}, \bibinfo {author} {\bibfnamefont {L.~W.}\ \bibnamefont {Anderson}}, \bibinfo {author} {\bibfnamefont {A.}~\bibnamefont {Agarwal}}, \bibinfo {author} {\bibfnamefont {L.~P.}\ \bibnamefont {Lindoy}}, \bibinfo {author} {\bibfnamefont {I.}~\bibnamefont {Rungger}}, \bibinfo {author} {\bibfnamefont {S.}~\bibnamefont {Mensa}},\ and\ \bibinfo {author} {\bibfnamefont {J.}~\bibnamefont {Crain}},\ }\bibfield  {title} {\bibinfo {title} {{Variational preparation of normal matrix product states on quantum computers}},\ }\href {https://arxiv.org/abs/2503.09683} {\bibfield  {journal} {\bibinfo  {journal} {arXiv}\ ,\ \bibinfo {pages} {2503.09683}} (\bibinfo {year} {2025})}\BibitemShut {NoStop}%
\end{thebibliography}%

\end{document}